# Micro- and Macrorheological Properties of Isotropically Cross-linked Actin Networks

Yuxia Luan, Oliver Lieleg, Bernd Wagner, Andreas R. Bausch
Lehrstuhl für Biophysik E22, Technische Universität München,
James-Franck-Straße 1, 85748 Garching, Germany

**ABSTRACT**

Cells make use of semi-flexible biopolymers such as actin or intermediate filaments to control their local viscoelastic response by dynamically adjusting the concentration and type of cross-linker molecules. The microstructure of the resulting networks mainly determines their mechanical properties. It remains an important challenge to relate structural transitions to both the molecular properties of the cross-linking molecules and the mechanical response of the network. This can be achieved best by well-defined *in vitro* model systems in combination with microscopic techniques. Here, we show that with increasing concentrations of the cross-linker HMM (heavy meromyosin) a transition in the mechanical network response occurs. At low cross-linker densities the network elasticity is dominated by the entanglement length $l_e$ of the polymer, while at high HMM densities the cross-linker distance $l_c$ determines the elastic behavior. Using microrheology the formation of heterogeneous networks is observed at low cross-linker concentrations. Micro- and macrorheology both report the same transition to a homogeneous cross-linked phase. This transition is set by a constant average cross-linker distance $l_c \approx 15$ µm. Thus, the micro- and macromechanical properties of isotropically cross-linked *in vitro* actin networks are determined by only one intrinsic network parameter.

**INTRODUCTION**

The proper functioning of living cells relies on the tight control of their local viscoelastic properties. Cells achieve this by the dynamic regulation of their cytoskeletal structures. To obtain a quantitative physical understanding of how the network microstructure correlates with the viscoelastic response, *in vitro* reconstituted networks have been proven to be crucial (for a recent review, see e.g. (1) and references therein). In this respect, the semi-flexible polymer actin has drawn much attention in recent years. The elastic response of solutions of actin filaments can be best understood in terms of confinement by surrounding polymers (2, 3). In contrast, the mechanics of cross-linked networks seem to depend much more on single polymer properties (4, 5). Moreover, depending on the molecular structure of the cross-linking molecules, different actin networks are formed. Actin binding proteins (ABPs) such as filamins (6, 7) or scruin (8) organize actin filaments into composite phases, while the ABPs fascin (5) and espin (9) as well as depletion forces (10, 11) enable the formation of purely bundled networks. Only with HMM in the rigor state a purely isotropically cross-linked network can be obtained for a broad range of ABP concentrations which allows theoretical predictions to be thoroughly tested (12). The static linear macroscopic response of isotropically cross-linked networks can be fully explained by a single filament model (4), while the frequency dependence of the loss modulus is determined by the unbinding kinetics of the cross-linkers (12). In these networks the non-linear elastic response is highly non-universal and dominated by unbinding events of cross-linking molecules. The linear elastic network response is determined by only one length scale, which is the average distance between cross-linkers $l_c$. Thus, homogeneously cross-linked actin networks are ideally suited to test the sensitivity of microrheological techniques on characteristic network length scales (13, 14).

Here, we show that the viscoelastic response of isotropically cross-linked actin networks is determined by the intrinsic network parameter $l_c$, both on the macroscopic and the microscopic scale. Two distinct regimes in the mechanical response can be distinguished: For very low HMM concentrations, the confinement of single filaments by neighboring polymers is sufficient to rationalize the obtained scaling behavior of the plateau modulus $G_0$ as a function of HMM concentration. At high cross-linker densities the affine stretching of single filaments dominates the linear elastic network response. The transition between both regimes is set by the cross-linker distance $l_c \approx 15$ μm, which is on the order of the persistence length of actin filaments. Cross-linker induced heterogeneities are identified for the low HMM concentration regime using microrheology. With increasing cross-linker concentration the relative network heterogeneity decreases until the critical cross-linker distance $l_c \approx 15$ μm is reached. At even higher HMM concentrations the relative distribution width of the local elastic modulus remains constant. With decreasing $l_c$ micro- and macrorheological techniques give converging results even in absolute numbers underlining the length dependence of microrheology.

**MATERIALS AND METHODS**

*Proteins*

G-actin is obtained from rabbit skeletal muscle and stored in lyophilized form at -21 °C (15). G-actin solution is prepared by dissolving lyophilized actin in deionized water and dialyzed against G-buffer (2 mM Tris, 0.2 mM CaCl$_2$, 0.2 mM DTT and 0.005 % NaN$_3$; pH 8.0) at 4 °C. The G-actin solution is kept at 4 °C and used within seven days. Gelsolin from bovine plasma serum is used to adjust the average filament length to 21 μm (16). Polymerization is initiated by adding 1/10 volume of 10×F-buffer (20 mM Tris, 20 mM MgCl$_2$, 2 mM CaCl$_2$, 1 M KCl and 2 mM DTT, pH 7.5) and gently mixing for 10 seconds. HMM is prepared from myosin II obtained from rabbit skeletal muscle following (17, 18). Its functionality was tested using standard motility assays (19). Actin networks with distinct amounts of HMM (heavy meromyosin) are investigated, tuning the molar ratio $R = \dfrac{c_{HMM}}{c_{actin}}$ over three decades from 0 to 1/50. For all samples the time dependence of the moduli is observed prior to measurements. This ensures that all ATP is depleted and HMM motor activity is arrested in its rigor state.

*Macrorheology*

The viscoelastic response of actin/HMM networks is determined by measuring the frequency-dependent viscoelastic moduli $G'(\omega)$ and $G''(\omega)$ with a stress controlled rheometer (Physica MCR 301, Anton Paar, Graz, Austria) within a frequency range of three decades. Approximately 500 μl sample volume is loaded within 1 min into the rheometer using a 50 mm plate-plate geometry with 160 μm plate separation. To ensure linear response only small torques (~ 0.5 μNm) are applied. Actin polymerization is carried out in situ, measurements are taken 60 min after the polymerization was initiated.

*Microrheology*

A magnetic tweezer microrheometer equipped with contrast microscopy is used to obtain local information on the viscoelastic moduli (20, 21, 22). The maximum force applicable with our setup is 5 pN, limiting the use of this technique to materials softer than ≈ 10 Pa. The viscoelastic moduli at a frequency of 0.1 Hz are determined at various positions in the sample, frequency spectra are taken within a frequency range of 2 - 3 decades depending on the sample stiffness. Approximately 20 μl sample volume is loaded into a cuvette, covered with a glass slide and sealed with vacuum grease. While the position of the magnetic coils is fixed with respect to the objective, the cuvette holder can be displaced. This ensures that all observed particles are placed in the centre of the magnetic coils. This facilitates the microrheometer calibration which is conducted in glycerol saturated with CsCl. Monodisperse paramagnetic beads (4.5 μm in diameter, Dynal M-450, Invitrogen, Karlsruhe, Germany) are added before polymerization and used as probing particles.

**RESULTS AND DISCUSSION**

In the rigor state HMM organizes an entangled actin solution into an isotropically cross-linked network (12). At low concentrations of HMM, ($R = \frac{c_{HMM}}{c_{actin}}$ < 1/2000), the macroscopic viscoelastic network response is similar to that of an entangled actin solution. Thus the elastic response can be understood in terms of deformation of reptation tubes. The frequency spectrum of the viscoelastic moduli is highly similar to the frequency behavior of an entangled actin solution (Fig. 1). Above a critical concentration $R^*_{macro}$ ($R > 1/2000$) the shape of the frequency spectra is drastically changed and the elastic response becomes very sensitively dependent on the HMM concentration: a pronounced plateau appears in the storage modulus $G'$ while the loss modulus $G''$ exhibits a minimum at low frequencies. This minimum in $G''$ is a signature of unbinding events of single cross-linkers occurring in the network (12). However, a minimum in $G''$ is not observable at low $R$ within the frequency range probed.

To quantify this mechanical transition at $R^*_{macro}$ the plateau modulus $G_0$ is determined at the minimum position in the loss modulus for distinct HMM concentrations. This is done for two different actin concentrations ($c_a$ = 0.4 mg/ml and $c_a$ = 0.8 mg/ml) as depicted in Fig. 2. Two distinct regimes of mechanical responses are observed: in the first regime, $G_0$ is almost independent of the HMM concentration, showing a scaling of $G_0 \sim R^{0.1}$. The observed dependence on the polymer density, $G_0 \sim c_a^{1.3}$, is in good agreement with predictions from the tube model for entangled actin solutions (2). Thus, at very low HMM densities a theoretical description based on the tube model holds – similar to observations for networks cross-linked by other actin-binding proteins as e.g. fascin (5) or α-actinin (23). In the second regime, $G_0$ is strongly dependent on the HMM concentration: $G_0 \sim R^{1.2}$. This agrees well with former findings on densely cross-linked HMM-networks ($R > 1/200$) (12). Moreover, also the dependence on the actin concentration as predicted by an affine stretching model (12, 4) is reproduced, $G_0 \sim c_a^{2.5}$. These different scaling regimes cross over at a critical molar ratio $R^*_{macro}$ which is only slightly dependent on the actin concentration ($R^*_{macro} = 1/2000$ for $c_a = 0.4$ mg/ml and $R^*_{macro} = 1/5000$ for $c_a = 0.8$ mg/ml).

To shed some light on the micromechanical origin of the transition in the macroscopic network response, local measurements of the network elasticity are performed using active microrheology (20). Qualitatively, the $G'$ and $G''$ frequency spectra obtained by microrheology exhibit the same transition in the overall shape as observed in macrorheology (Fig. 3). The transition point $R^*_{micro} = 1/2000$ obtained from microrheology agrees very well with the mechanical transition point $R^*_{macro}$ determined by macrorheology. This suggests that the observed mechanical transition is guided by an intrinsic network length scale on the order of several microns as microrheology is sensitive to such small length scales.

For a detailed statistical analysis of putative local heterogeneities, the viscoelastic moduli are determined at various positions in the sample (approximately 100 positions per sample) for an intermediate frequency (0.1 Hz). From the distribution data of $G'(0.1$ Hz$)$ cumulative probabilities are calculated for different $R$ (Fig. 4 (A)). Also in the form of the distributions two regimes can be distinguished: below $R^*_{micro}$ the distribution curves show the occurrence of pronounced heterogeneities while above the critical concentration $R^*_{micro} = 1/2000$ homogeneous distributions are obtained. In the latter, the mean values and the absolute widths of the distributions increase with increasing cross-linker concentration $R$. Thus, normalizing the cumulative probabilities by their respective average elastic modulus (Fig. 4(B)) results in a collapse onto a single curve. Interestingly, this normalized distribution curve is identical with the distribution curve obtained for an entangled actin solution, nicely confirming the homogeneous and isotropic cross-linking effect of rigor-HMM (12).

The relative distribution width is given by the normalized standard deviation $\sigma_{rel} = \sigma / <G'(0.1Hz)>$. A relative width of $\sigma_{rel} \approx 20$ % is obtained for an entangled actin solution as well as for cross-linked networks ($R > R^*_{micro}$) as depicted in the inset of Fig. 4 (B). At low HMM concentrations ($R < R^*_{micro}$), $\sigma_{rel}$ is more than twice as large corresponding to the broad shape of the distribution curves: the degree of heterogeneity is increased by the addition of a few cross-linking molecules, until at the transition point $R^*_{micro}$ a homogenous microstructure is reached again.

The $G'$ distributions obtained for the two lowest ratios ($R = 1/5000$ and $R = 1/10000$) contain values even lower than the elastic modulus of an entangled actin solution; the corresponding $G'$ histogram exhibits a second peak at $G'(0.1$ Hz$) \approx 20$ mPa. At the same time, only very few values exceed the elastic modulus obtained for $R = 0$. To elucidate the network microstructure responsible for this peculiar broadening, full frequency spectra are determined at different sample positions. All obtained spectra have comparable shapes, none of them exhibits a minimum in $G''(\omega)$ within the frequency range probed. As a minimum in $G''(\omega)$ would be a signature of a cross-linked network, this suggests that in this concentration regime the viscoelastic response is determined rather by local filament density fluctuations induced by the few HMM molecules present than by a formation of cross-linked microdomains, as suggested for α-actinin (23).

To elucidate the important network length scales which are responsible for the distinct mechanical scaling regimes we compare the macrorheological results with the local magnetic tweezer measurements. The frequency behavior of $G'$ and $G''$ is determined at distinct positions in the cross-linked network using active microrheology. The resulting frequency spectra are averaged to obtain a quantity that represents the frequency dependent response of the whole network – although locally measured. $G_0$ is determined as described before and compared to the macroscopic results as depicted in Fig. 5. The same $R^*$ is obtained from these two techniques. While below $R^*$ the microscopic modulus is approximately a factor of 3 - 4 lower compared to the macroscopic value, both measurements agree better and better with raising $R$. Finally, both curves merge at $R^\# = 1/100$ as can clearly be seen from the inset of Fig. 5 which depicts the direct ratio $\varsigma = G_{0,macro}/G_{0,micro}$ on dependence of $R$. It is important to note, that for the low $R$ samples showing the broad distributions in the microscopic modulus the disagreement between the microrheological and the macrorheological measurements is most pronounced, $\varsigma \approx 4$. While for the entangled actin solution
$\varsigma \approx 2.5$, this ratio decreases down to $\varsigma \approx 1$ at $R^{\#} \approx 1/100$. Beyond $R^{\#}$ both techniques report the same absolute values for the plateau modulus $G_0$, indicating that both moduli obtained by the micro- and macrorheologial techniques are determined by the same mesoscopic length scale.

For a semi-flexible polymer network the mechanical response crucially depends on the length scale probed. Therefore, it would be desirable to correlate both experimentally determined parameters $R^*$ and $R^{\#}$ to intrinsic network length scales. For an isotropically cross-linked network the dominating length scale is the cross-linker distance $l_c$, which can be determined by the onset of the non-linear regime. The elasticity of the network should, with decreasing $l_c$, become non-linear at large strains. Non-linear effects will set in at a critical deformation $\gamma_{crit}$ from which $l_c$ can be calculated. For isotropically cross-linked HMM networks $l_c$ is given as $l_c = 1.6 \cdot l_p \cdot \gamma_{crit}$ (12). Therefore, the transition ratio $R^*$ can be physically understood: the crossing-over from one scaling regime to the other occurs at well defined $l_c$, which is approximately constant within the error bars (17 ± 2 μm for $c_a$ = 0.4 mg/ml, 14 ± 2 μm for $c_a$ = 0.8 mg/ml). When $G_0$ is parameterized in the two distinct scaling regimes (5), a constraint relating $R^*_{macro}$ to the actin concentration is obtained, as $R^*_{macro} \sim c_a^{-6/5}$. Introducing the entanglement length $l_e \sim c_a^{-2/5} \cdot l_p^{1/5}$ results in the relation: $c_{HMM} \cdot l_e^{-1/2} \approx const$. Thus, with increasing actin concentration (corresponding to decreasing $l_e$) smaller amounts of HMM molecules are necessary to induce the crossing-over to the second scaling regime. Thus, the critical parameter which is held constant along the "phase boundary" is nothing else than the average cross-linker distance $l_c$ as no overall rearrangement in the network structure occurs. One might speculate that $l_c \sim$ 14-17 μm could refer to the percolation threshold at which the cross-link density is high enough to evoke an overall network response that is not dominated by entanglements any more. This in turn is set by the persistence length – determining the length scale for tube confinement effects. As in the case of HMM networks the overall network structure will not be changed when crossing $R^*_{macro}$, no extra free energy or enthalpy compensations is needed for structural rearrangements and less HMM molecules are required to induce the mechanical transition at higher filament densities. This stands in marked contrast to what was observed for the bundling transition reported for fascin networks (5). In the latter case the formation of a purely bundled phase will lead to a strong decrease in configurational entropy, which has to be balanced with a sufficient gain in binding enthalpy. Thus, for the bundling transition of fascin networks a qualitative different constraint was obtained, $c_{fascin} \cdot l_e^{+1/2} \approx const$.

The parameter $R^{\#}$, where both the microscopic and the macroscopic measurement technique report the same viscoelastic response in absolute numbers, corresponds to a cross-linker distance of $l_c \approx 5$ μm. This excellently matches the diameter of the beads used for active microrheology. At this point the bead explores the full spectrum of dominating relaxation modes while at larger $l_c$ (corresponding to lower $R$) long-wavelength fluctuations are not fully reported (13). Thus, the plateau modulus will be underestimated, resulting in the observed discrepancy of the microrheological and macrorheological measurements.

Still, there remains a small difference between the macroscopic and the microscopic frequency spectrum, even at $R > R^{\#}$, as the minimum in $G''(\omega)$ always occurs at lower frequencies in the case of microrheology. At the same time, the uprise of $G''(\omega)$ at even smaller frequencies is more pronounced in the microscopic frequency spectrum. This is attributable to the fact that a minimum in $G''(\omega)$ is a signature of unbinding events occurring at cross-linking points. As the microrheological technique used in this study is much more sensitive to local changes in the microstructure, unbinding events in the vicinity of the probing particle will be reported much easier using a magnetic tweezer than a macrorheometer.

In conclusion, we have shown that combining active micro- and macrorheology is a powerful tool to elucidate the structure-mechanics relation of semi-flexible polymer networks. The purely isotropically cross-linked network obtained by adding rigor-HMM to actin solutions has proven to be a powerful tool to obtain not only a microscopic understanding of the mechanics but also a detailed description of the structural transition induced by the cross-linker molecules. The induced heterogeneities at low cross-linker concentrations may be a more generic phenomenon and a prerequisite for the main structural transition into a network structure dominated by cross-links. The here obtained physical description of the transition between entangled to isotropically cross-linked networks will be indispensable for the more complicated cross-linker induced transitions into composite or purely bundled phases.

We thank M. Rusp for protein preparation. Financial support from the Volkswagen Foundation and the German Excellence Initiatives via the "Nanosystems Initiative Munich (NIM)" is gratefully acknowledged. O. Lieleg acknowledges support from CompInt in the framework of the ENB Bayern.

Figure legends:

FIGURE 1 Frequency dependence of the viscoelastic moduli for different concentrations of HMM at a constant actin concentration ($c_{actin}$ = 0.4 mg/ml) as determined by macrorheology. With increasing HMM concentration a pronounced plateau in the storage modulus occurs above $R = 1/200$, which is accompanied by a deep minimum in the loss modulus.

FIGURE 2 Plateau modulus $G_0$ as a function of the molar ratio $R$ as determined by macrorheology for two different actin concentrations ($c_a$ = 0.4 mg/ml and $c_a$ = 0.8 mg/ml). For both actin concentrations measured two different scaling regimes are observed as described in the text.

FIGURE 3 Frequency dependence of the viscoelastic moduli for different $R$ as determined by microrheology. The frequency spectra exhibit the same transition in their overall shape as also reported from macrorheology. However, the minimum in $G''$ is shifted to lower frequencies and more pronounced than that in the case of the macrorheological measurement.

FIGURE 4 Cumulative probabilities are obtained from the local distribution data of $G'$ for different $R$ from approximately 100 beads per sample (A). Normalization by the respective average value $<G'>$ allows an overlay of curves for $R > R^*_{micro} = 1/2000$ as shown in (B). This is not possible for samples $R < R^*_{micro}$ as their relative distribution width $\sigma_{rel} = \sigma/<G'>$ is almost twice as large as for the other samples as depicted in the inset.

FIGURE 5 Plateau modulus $G_0$ as obtained from microrheology (squares) and macrorheology (circles) as a function of $R$. The ratio of the plateau moduli $\varsigma = G_{0,macro}/G_{0,micro}$ is depicted in the inset. Above $R = 1/100$ (corresponding to $l_c \approx 5$ μm) micro- and macrorheology report the same absolute values.

**Figure 1**

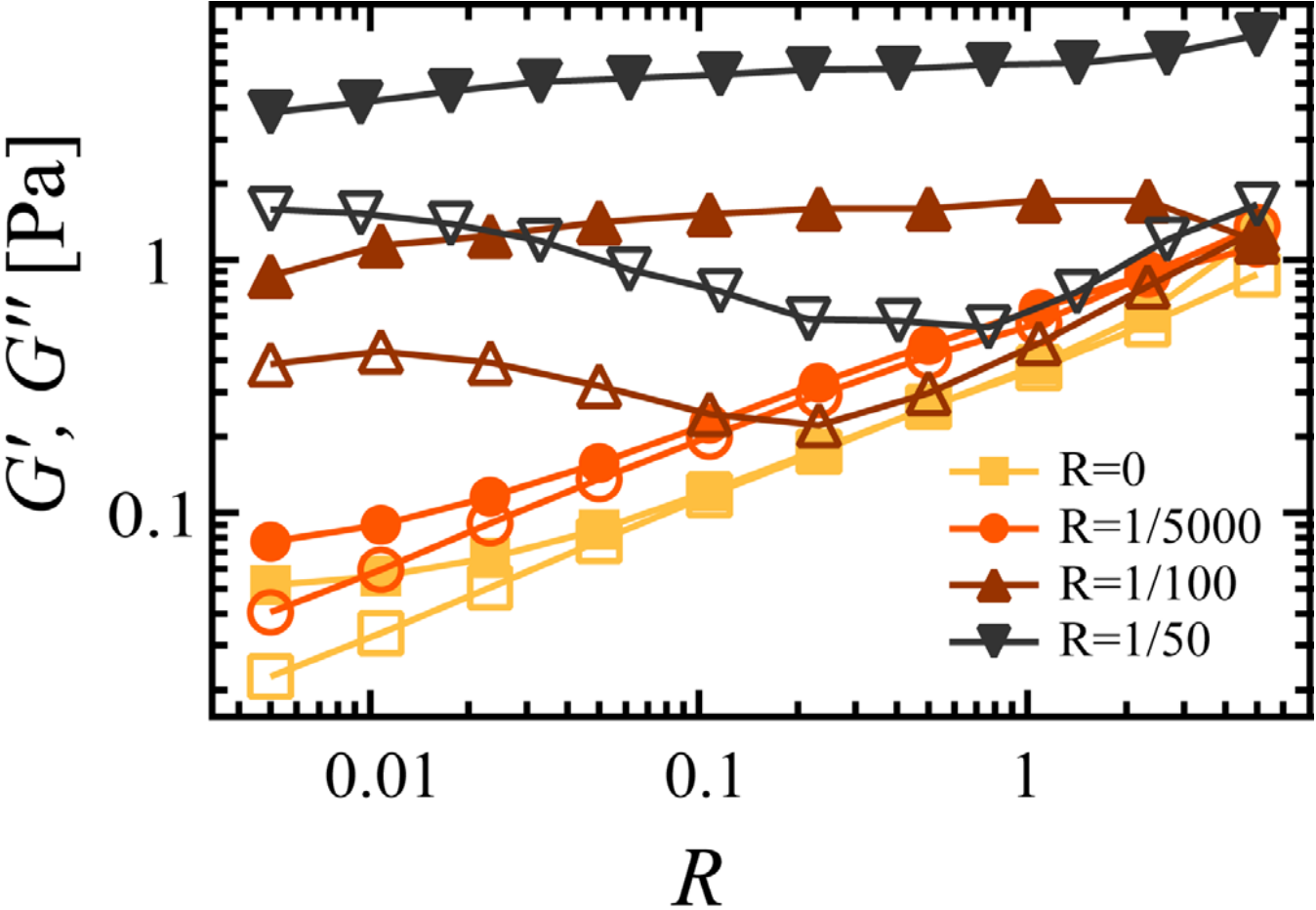

**Figure 2**

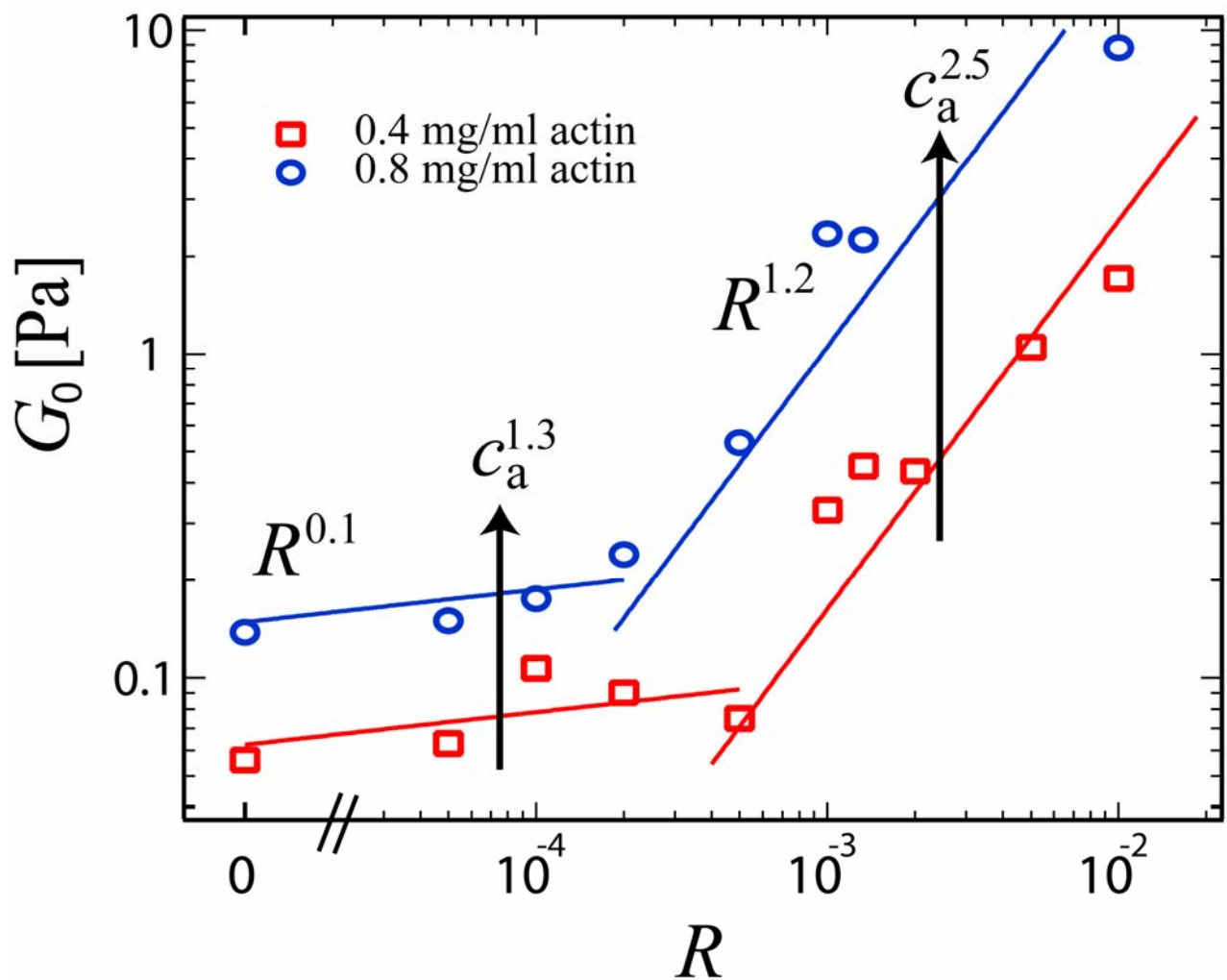

**Figure 3**

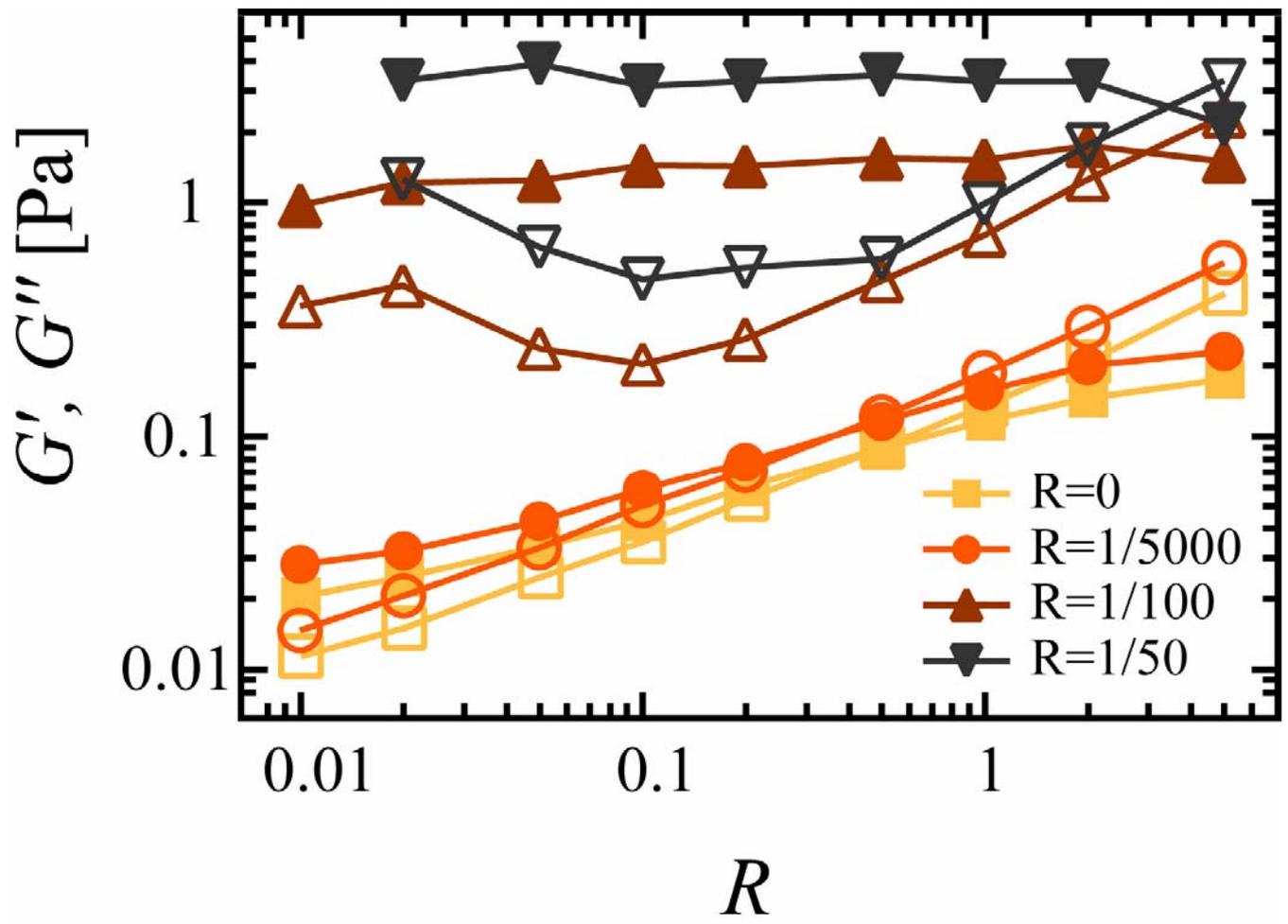

**Figure 4**

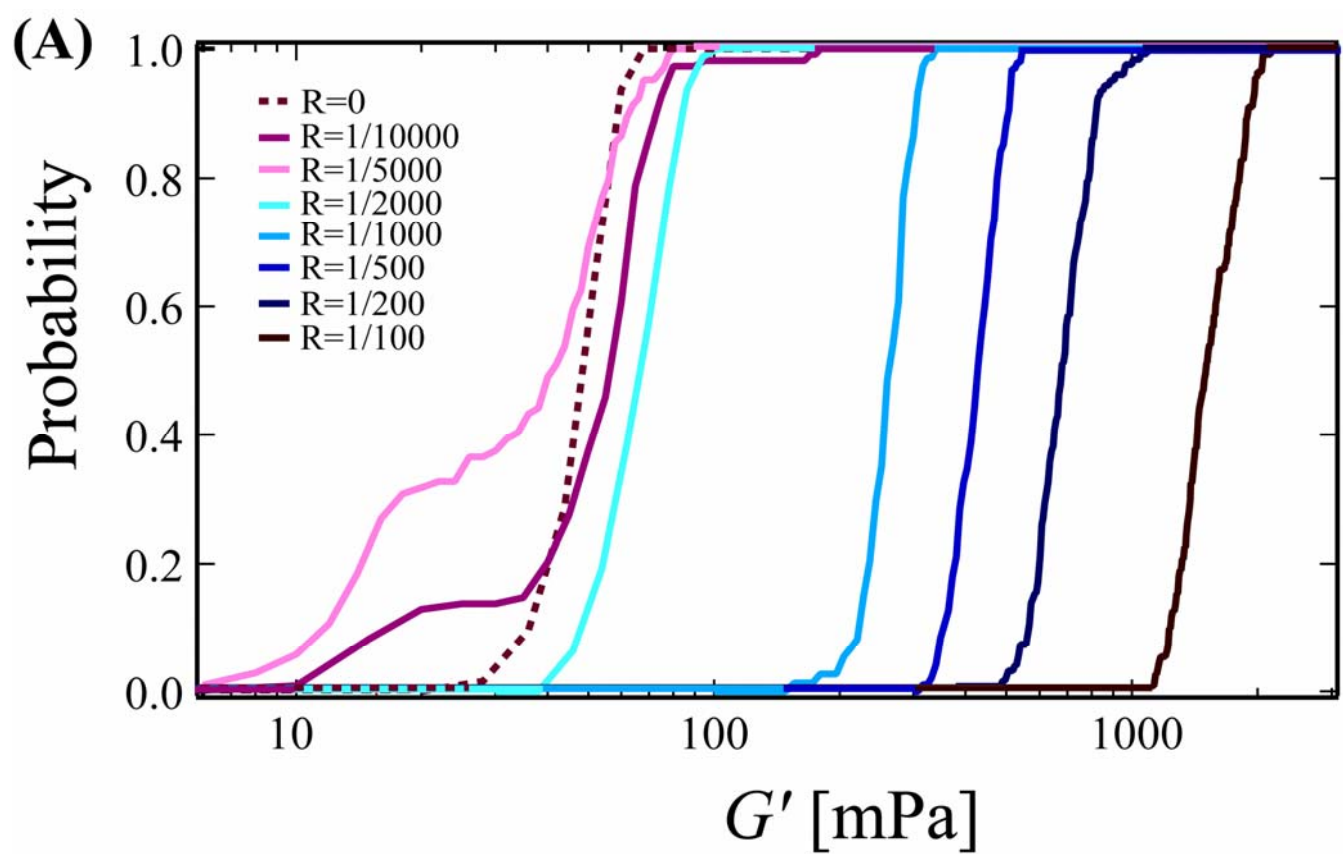

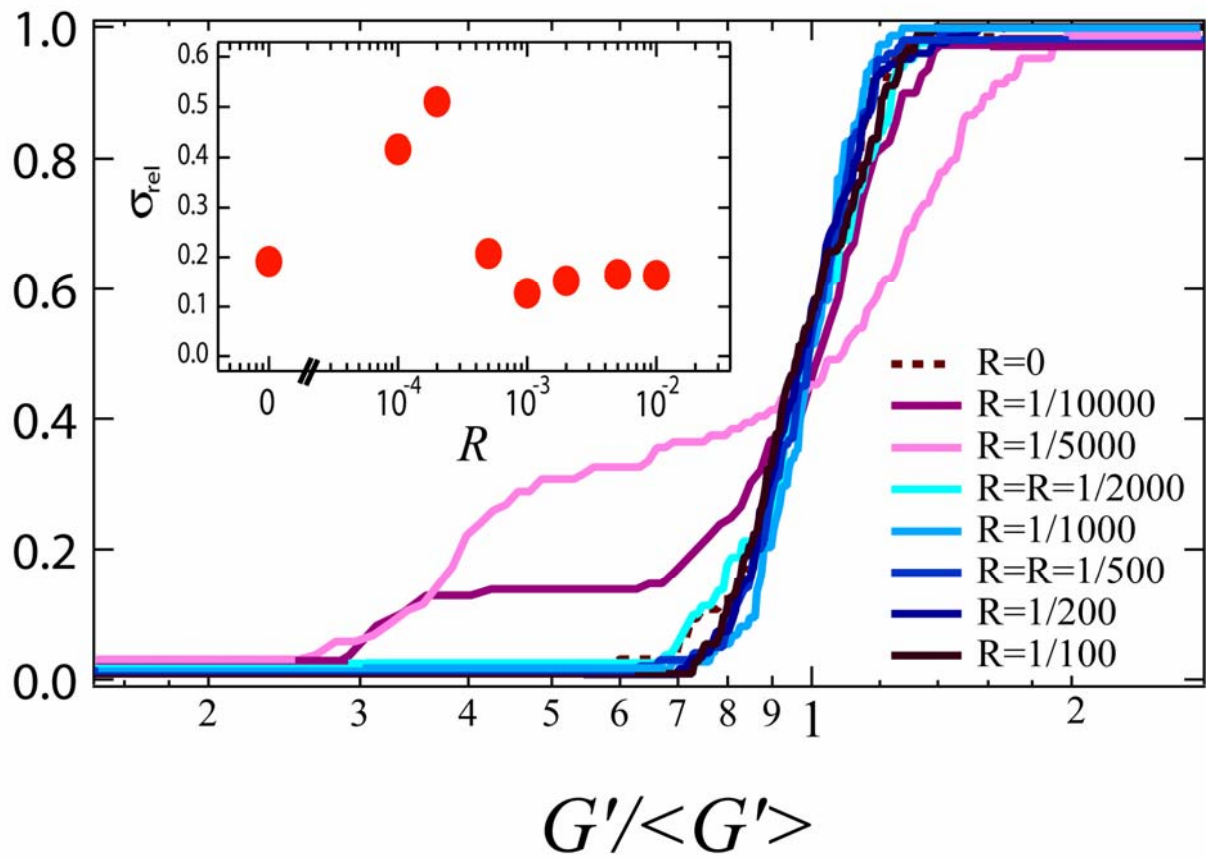

**Figure 5**

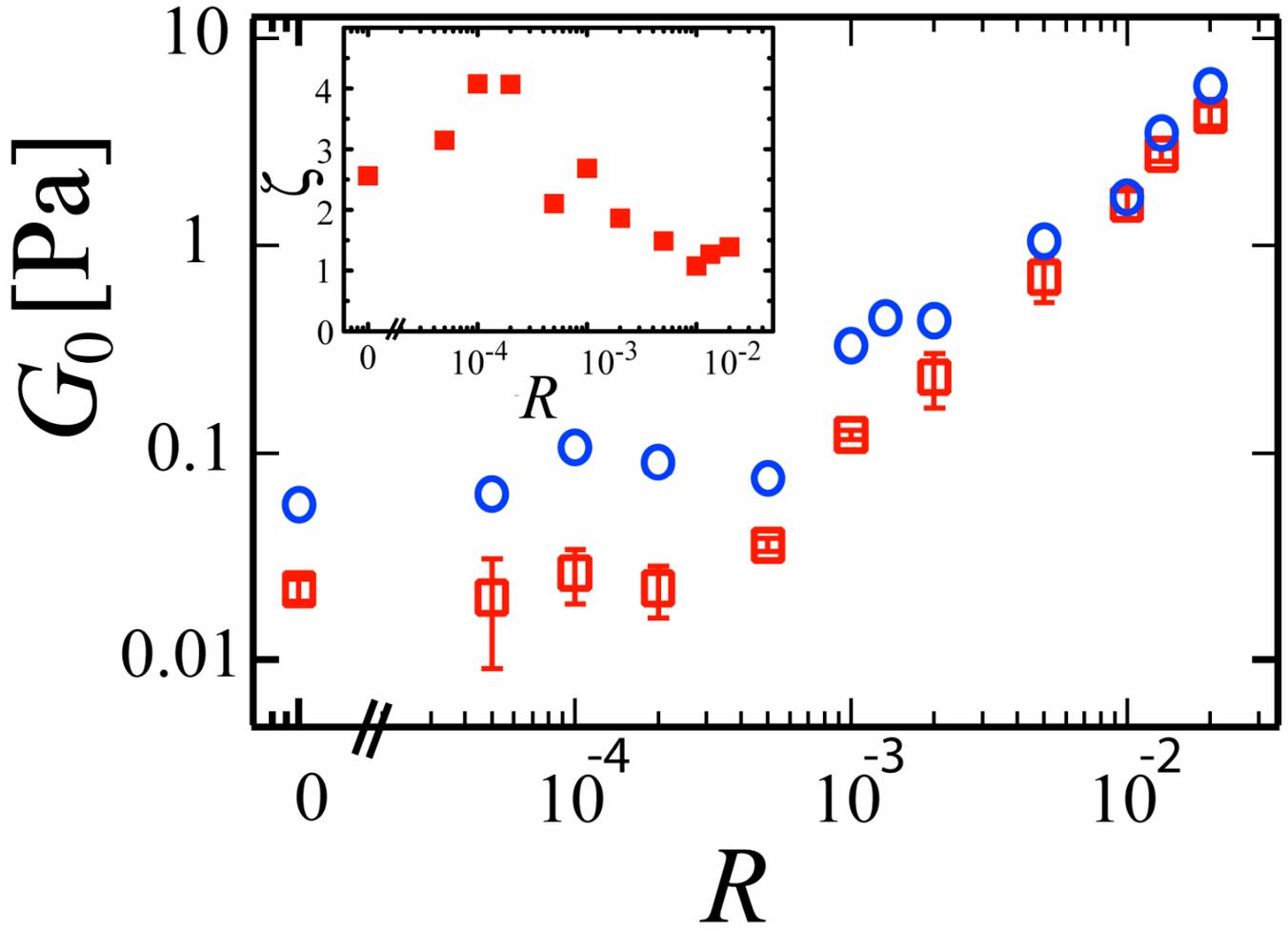